\documentclass[aps,twocolumn,prd]{revtex4}
\usepackage{amsmath,amssymb,epsfig,natbib,bm}
\usepackage{hyperref}
\usepackage{subfigure}
\newcommand{\be}{\begin{equation}}
\newcommand{\ee}{\end{equation}}
\newcommand{\ba}{\begin{eqnarray}}
\newcommand{\ea}{\end{eqnarray}}
\newcommand{\cg}[6]{\mathcal{C}^{#1 #2}_{#3 #4 #5 #6}}
\newcommand{\wninej}[9]{\left \{
                           \begin{array}{ccc}
        \! #1\! & #2\!  & #3\!  \\
        \! #4\! & #5\!  & #6\!  \\
	\! #7\! & #8\!  & #9\!
                           \end{array}
                   \right \} }
\newcommand{\wsixj}[6]{\left\{
                           \begin{array}{ccc}
         #1 & #2  & #3  \\
         #4 & #5  & #6
                           \end{array}
                   \right\}}
\begin{document}
\title{Statistical Isotropy violation of the CMB brightness fluctuations}
\author{Moumita Aich}
\email{moumita@iucaa.ernet.in}
\author{Tarun Souradeep}
\email{tarun@iucaa.ernet.in}
\affiliation{Inter-University Centre for Astronomy and Astrophysics, Post Bag 4, Ganeshkhind, Pune 411 007, India.}
\begin{abstract}
Certain anomalies at large angular scales in the cosmic microwave
background measured by WMAP have been suggested as possible evidence
of breakdown of statistical isotropy(SI). SI violation of cosmological
perturbations is a generic feature of ultra large scale structure of
the cosmos and breakdown of global symmetries. Most CMB photons
free-stream to the present from the surface of last scattering. It is
thus reasonable to expect statistical isotropy violation in the CMB
photon distribution observed now to have originated from SI violation
in the baryon-photon fluid at last scattering, in addition to
anisotropy of the primordial power spectrum studied earlier in
literature.

We consider the generalized anisotropic brightness distribution
fluctuations, $\Delta(\vec{k}, \hat{n}, \tau)$ (at conformal time
$\tau$) in contrast to the SI case where it is simply a function of
$|\vec{k}|$ and $\hat{k} \cdot \hat{n}$.  The brightness fluctuations
expanded in Bipolar Spherical Harmonic (BipoSH) series, can then be
written as $\Delta_{\ell_1 \ell_2}^{L M}(k, \tau)$ where $L > 0
$ terms encode deviations from statistical isotropy. Violation of SI
encoded in the present off-diagonal elements of the harmonic space
correlation $\langle a_{\ell m} a_{\ell' m'}^* \rangle$, equivalently,
the BipoSH coefficients $A^{LM}_{\ell \ell'}$, are then related to
the generalized BipoSH brightness fluctuation terms at present.  We
study the evolution of $\Delta_{\ell_1 \ell_2}^{L M}(k, \tau)$
from non-zero terms $\Delta_{\ell_3 \ell_4}^{L M}(k, \tau_s)$ at
last scattering, in the free streaming regime. We show that
the terms with given BipoSH multipole, $LM$, evolve
independently. Moreover, similar to the SI case, power at small
spherical harmonic (SH) multipoles of $\Delta_{\ell_3 \ell_4}^{L M}(k,
\tau_s)$ at the last scattering, is transferred to $\Delta_{\ell_1
\ell_2}^{L M}(k, \tau)$ at larger SH multipoles.  The structural
similarity is more apparent in the asymptotic expression for large
values of the final SH multipoles. \emph{This formalism allows an elegant
identification of any SI violation observed today to a possible origin
in SI violating physics present in the baryon-photon fluid.} This is illustrated 
for the known result of SI violating angular correlations due to the presence of 
a homogeneous magnetic field in the baryon-photon fluid. 

\end{abstract}
\maketitle
\section{Introduction}
The Cosmic Microwave Background (CMB) anisotropy is a very powerful observational probe of cosmology.
In standard cosmology, CMB anisotropy signal is expected to be statistically isotropic, i.e., statistical expectation
values of the temperature fluctuations $\Delta T(\hat{n})=\sum_{\ell m} a_{\ell m} Y_{\ell m}(\hat{n})$ are preserved under rotations of the sky.
The condition for statistical isotropy (SI), in spherical harmonic space translates to a diagonal $\langle a_{\ell m} a_{\ell' m'}^* \rangle = C_{\ell} \delta_{\ell \ell'}
\delta_{m m'}$ where $C_{\ell}$ is the widely used angular power spectrum of the CMB anisotropy.

\noindent After the release of first year data of the Wilkinson Microwave Anisotropy Probe (WMAP), statistical isotropy of the CMB anisotropy
attracted considerable attention. The study of full sky maps from the WMAP 5 year data \cite{komatsu,spergel,nolta} and the very recent WMAP 7 year data
\cite{bennett}, has led to some intriguing anomalies which seem to suggest that the assumption of statistical isotropy is broken on
the largest angular scales \cite{costa,copi,hansen,eriksen,hoftuft}.
Broken isotropy would have a profound implications for standard cosmological model as statistical
isotropy underlies all cosmological inferences.

\noindent It was pointed out that the suppression of power in the quadrupole and
octopole are aligned in the form of the `` axis of evil'' \cite{land,tegmark,magueijo,rakic,frommert}.
Further ``multipole-vector" directions associated
with these multipoles (and some other low multipoles as well) appear to be anomalously correlated \cite{copi}, \cite{copi04}, \cite{schwarz}.
There are indications of asymmetry in the power spectrum at low multipoles in opposite hemispheres, the
``north-south asymmetry'' \cite{hansen}, \cite{eriksen}, \cite{eriksen0,hansen1,naselsky}. Possibly related, are the results
of tests of Gaussianity that show asymmetry in the amplitude of the measured genus amplitude (at about 2 to 3 $\sigma$ significance)
between the north and south galactic hemispheres \cite{park,eriksen1,eriksen2}. Analysis of the distribution of extrema in
WMAP sky maps has indicated non-Gaussianity, and to some extent, violation of SI \cite{wandelt}.

\noindent An observed map of CMB anisotropy, $\Delta T(\hat{n})^{\mathrm{obs}}$ contains the true CMB temperature $\Delta T(\hat{n})$ fluctuations,
convolved with the beam and instrumental noise \& foreground contaminations. Breakdown of statistical isotropy can occur in any of
these parts and can be categorized as

\begin{itemize}

\item Theoretically motivated effects which are intrinsic to the true CMB sky, $\Delta T(\hat{n})$ include non-trivial cosmic topology \cite{ts},
Bianchi models \cite{jaffe05,jaffe06,pontzen,tuhin} and primordial magnetic fields \cite{durrer}, \cite{kahniashvili}.
A recent article \cite{urban} claims that the solution to the cosmological vacuum energy can be explained as a result of the interaction of
the infrared sector of the effective theory of gravity with standard model fields. This theory predicts the violation of cosmological isotropy.

\item Although a possible source of SI breakdown, residual foreground contamination would need to be of order the intrinsic
CMB temperature anisotropy to account for an appreciable effect \cite{copi06,hinshaw,gold}.

\item It would be erroneous to assume that the true CMB temperature fluctuations are completely extracted
from the observed map. Observational artifacts such as non-circular beam, inhomogeneous noise correlation,
residual striping patterns could be potential sources of SI breakdown.

\end{itemize}

Violation of statistical isotropy of CMB anisotropy and its measurement has been discussed in literature earlier 
\cite{hajian-2003,hajian,souradeep-2003,hajian-2004,hajian-2006,souradeep}
by defining an estimator where SI breakdown in an observed CMB anisotropy sky map is indicated by non zero value of this estimator.
Studies have also been done by implementing a directional dependent inflationary power spectrum $P(\vec{k})$ which gives rise to off-diagonal
terms in the covariance matrix \cite{ackerman}, \cite{pullen}; spontaneous breakdown of SI in the CMB by a non-linear response to long-wavelength
field fluctuations that appear as a gradient locally to the observer \cite{gordon} or locally through a modulation field \cite{dvorkin};
incorporating an initial period of kinetic energy domination in single field inflation \cite{donoghue}. 

In this paper we present a new formulation that relates the breakdown of SI in the CMB photon
fluctuations at last scattering, and evolving them to find the effect of the modes at present epoch hence the CMB 
$\langle a_{\ell m} a_{\ell' m'}\rangle$ today. We also find the Bipolar spherical harmonic coefficients (BipoSH) \cite{hajian-2004,hajian-2006} which are
linear combinations of off-diagonal elements of the covariance matrix. BipoSH expansion completely represents the information of the covariance matrix
thus being the most general way of studying two point correlation functions of CMB anisotropy.
These BipoSH coefficients are mathematically complete measures of SI violation on a sphere. 
\section{Review of Statistically Isotropic CMB brightness fluctuations}
\subsection{Boltzmann equations, inhomogeneities and anisotropies}

In the smooth background universe, thermalized photons being distributed homogeneously and isotropically, the
temperature $T$ is independent of $\vec{x}$ and direction of propagation $\hat{p}$ respectively. To describe perturbations about this smooth universe,
we allow inhomogeneities in the photon distribution and anisotropies. 

Before recombination, $z_{\mathrm{rec}} \approx 1100$, the photons were tightly coupled to the electrons and protons; all together they can be described 
as a single fluid, the baryon-photon fluid. After recombination, photons free-stream from the surface of last scattering to the present epoch.

Given the cosmological perturbations to the photons at recombination, one can predict the anisotropy spectrum today.
The main motivation next is to relate the moments today to the moments at recombination using the photon distribution function. 

\subsection{\label{sec:delta} Fluctuations of CMB photon distribution}

In the Boltzmann equation for photons $df/dt = C[f]$, we expand the photon distribution function
$f(\vec{x},p,\hat{n},\tau)$ about its zero-order Bose-Einstein value $T(\tau)$ \cite{dodelson} where $\vec{p} = p \hat{n}$. 
The distribution function of the photons changes with the perturbed temperature as
\be
f(\vec{x}, p, \hat{n}, \tau) = \left[ \mathrm{exp} \left \{ \frac{p}{T(\tau) [1+\Delta(\vec{x}, \hat{n}, \tau)]}  \right \} -1 \right] ^{-1}
\label{boltz}
.\ee
The perturbation to the distribution function is characterized by $\Delta \equiv \delta T/T$ termed as CMB brightness fluctuations henceforth.
Since the perturbation $\Delta$ is small, we can expand $f(\vec{x}, p, \hat{n}, \tau)$ keeping only terms up to first order to get
\be
\Delta(\vec{x},\hat{n},\tau) \equiv \left(\frac{\partial f^0}{\partial \ln p}\right)^{-1} \delta f ,
\label{pert}
\ee
where $f^0$ is the zero-order photon distribution function. 
$\Delta(\vec{x},\hat{n},\tau)$ depends on $\vec{x}, \hat{n}$ and $\tau$ and not on the magnitude of momentum $p$; this is a valid assumption since the temperature
of the plasma is very small compared to the rest energy of the electrons which undergo scattering, elastic Thomson scattering
has negligible effect on the magnitude of the photon momentum.

\vspace{2mm}

\noindent Perturbations to the CMB remain small at all cosmological epochs; evolution of the largest scales being in the linear regime.
In solving the linear evolution equations, it is simplest to work with Fourier transforms since every Fourier mode evolves independently.
\be
\Delta(\vec{x}, \hat{n}, \tau) = \int \frac{d^3 k}{(2\pi)^3} e^{i \vec{k} \cdot \vec{x}} \tilde{\Delta}(\vec{k}, \hat{n}, \tau) \phi(\vec{k}) \; ,
\label{fourier}
\ee
\noindent where $\phi(\vec{k})$ is the primordial density fluctuations.

\vspace{5mm}

\noindent With statistical isotropy assumption
\ba
\tilde{\Delta}(\vec{k}, \hat{n}, \tau) &=& \tilde{\Delta}(k, \hat{k} \cdot \hat{n}, \tau) \nonumber \\
&=&  \sum_\ell (-i)^\ell (2\ell +1)
\tilde{\Delta}_\ell(k,\tau) P_\ell(\hat{k} \cdot \hat{n}) \nonumber \\
&=&  4 \pi \sum_{\ell m} (-i)^{\ell} \tilde{\Delta}_\ell(k,\tau) Y_{\ell m}(\hat{k}) Y_{\ell m}(\hat{n})
\label{delta}
.\ea
$\tilde{\Delta}_\ell(k,\tau)$ are the moments of the CMB brightness fluctuation. The monopole $\tilde{\Delta}_0$ is related to the density perturbations
while the dipole $\tilde{\Delta}_1 \propto \hat{n} \cdot \vec{v}$, gives the velocity term for baryons.

\subsection{\label{sec:corr} Correlations}

The observed anisotropy in multipole space can be written in terms of the CMB brightness fluctuation as
\be
\Delta(\vec{x} = 0, \hat{n}, \tau) = \sum_{\ell m} a_{\ell m} Y_{\ell m}(\hat{n}) \; .
\label{obsanis}
\ee
Using the orthonormality property of the spherical harmonics $Y_{\ell m}$, the SH coefficients $a_{\ell m}$ become
\be
a_{\ell m} = 4 \pi (-i)^{\ell} \int \frac{d^3 k}{(2\pi)^3} \; \phi(\vec{k}) \tilde{\Delta}_\ell (k,\tau) Y_{\ell m}^*(\hat{k})
.\ee
The angular correlation can be expressed as
\ba
\langle a_{\ell m} a_{\ell' m'}^*\rangle &=& 4\pi \int \frac{dk}{k} \frac{k^3 P_0(k)}{2\pi^2} |\Delta_\ell(k,\tau)|^2
\delta_{\ell \ell'} \delta_{m m'}  \nonumber \\
C_{\ell} &=& 4\pi \int \frac{dk}{k} \mathcal P_0(k) |\Delta_\ell(k,\tau)|^2 \; ,
\label{corr}
\ea
\noindent where correlation of the primordial density fluctuations $\langle \phi(\vec{k})\phi^*(\vec{k'})\rangle = P_0(k) \delta(\vec{k} - \vec{k'})$ and 
$\mathcal P_0(k)=k^3 P_0(k)/2\pi^2 $ is the primordial power spectrum per logarithmic interval generated by inflationary model and the second term is the 
radiative transport kernel in the post-recombination universe given by cosmological parameters.
\subsection{\label{sec:evol} Evolution of CMB brightness fluctuation in the free-streaming regime}

The evolution of \! $\tilde{\Delta}(\vec{k},\hat{n},\tau)$ in the free-streaming regime can be written as
\be
\tilde{\Delta}(\vec{k}, \hat{n}, \tau)  = e^{i \vec{k} \cdot \hat{n} (\tau - \tau_s)} \tilde{\Delta}(\vec{k}, \hat{n}, \tau_s),
\label{free}
\ee
\noindent where $\tau$ is well inside the free-streaming regime i.e. $\tau_s < \tau < \tau_0$, $\tau_s$ and $\tau_0$ 
being the conformal time at last scattering and today respectively \cite{bond}.

\noindent Using the expansion 
\be
e^{i \vec{k} \cdot \hat{n}\Delta \tau}  = \sum_l (-i)^l (2l+1) j_l(k \Delta \tau) P_l(\hat{k} \cdot \hat{n})
\label{plane}
,\ee
and defining $\Delta \tau = \tau - \tau_s$, the evolution equation for $\tilde{\Delta}_\ell(k,\tau)$ can be written as
\ba
\tilde{\Delta}_{\ell}(k,\tau) &=& \sum_{l \ell'} (-i)^{\ell+\ell'-l} \: (2\ell'+1) \: j_l(k \Delta \tau) \nonumber \\
&\times& [\cg l 0 {\ell} 0 {\ell'} 0]^2 \: \tilde{\Delta}_{\ell'}(k,\tau_s)
\label{evolve}
.\ea
$j_l(k \Delta \tau)$ and $P_l(\hat{k} \cdot \hat{n})$ are the $\ell^{\mathrm{th}}$ order spherical Bessel function and Legendre Polynomial, respectively.
Equation (\ref{evolve}) is the well known ``free streaming '' equation in CMB literature \cite{bond}. 
$\mathcal{C}^{L M}_{\ell_1 m_1 \ell_2 m_2}$ are the Clebsch-Gordan coefficient which satisfies the triangle inequalities (\cite{varsha})
putting a constraint $|\ell_1-\ell_2| \leq L \leq \ell_1+\ell_2$ and $m_1+m_2 = M $.

\section{Statistical isotropy breakdown in the CMB brightness fluctuation}
In this paper we take into account the SI violation of the CMB anisotropy which is seeded due to the inherent SI breakdown
in the CMB photon distribution. We consider the general form of the CMB brightness fluctuation, allowing for anisotropy in $\hat{k}$, i.e.,
$\tilde{\Delta}(\vec{k}, \hat{n}, \tau) \not\equiv  \tilde{\Delta}(k, \hat{k} \cdot \hat{n}, \tau)$.

\subsection{Generalized CMB brightness fluctuations}
The most general CMB brightness fluctuation is not simply a function of $|\vec{k}|$ and $\hat{k} \cdot \hat{n}$.  In this case the
physical situation of anisotropic fluctuations demand the brightness fluctuations to be expanded in Bipolar Spherical Harmonic series
(not just a Legendre series as in the statistical isotropic case). The brightness fluctuation in multipole space is
$\Delta_{\ell_1 \ell_2}^{L M}(\vec{k}, \tau)$ where $L > 0 $ term incorporates deviation from statistical isotropy.

\ba
\tilde{\Delta}(\vec{k}, \hat{n}, \tau) &=& 4 \pi \sum_{\ell_1 \ell_2 L M} (i)^{\frac{\ell_1+\ell_2}{2}}
\sqrt{\ell_1 + \ell_2 + 1} \nonumber  \\
& \times & \tilde{\Delta}_{\ell_1 \ell_2} ^{L M}(k,\tau)  \; \{Y_{\ell_1} (\hat{k}) \otimes Y_{\ell_2} (\hat{n})\}_{LM}  \nonumber \\
&=& 4 \pi \sum_{\ell_1 \ell_2 L M} \beta_{\ell_1 \ell_2} \; \tilde{\Delta}_{\ell_1 \ell_2} ^{L M}(k,\tau) \nonumber \\
& \times& \{Y_{\ell_1} (\hat{k}) \otimes Y_{\ell_2} (\hat{n})\}_{LM} ,
\label{deltadef}
\ea
where $\beta_{\ell_1 \ell_2} = (i)^{\frac{\ell_1+\ell_2}{2}} \sqrt{\ell_1 + \ell_2 + 1}$ 
has been defined for convenient notational simplicity.

\noindent The tensor product in BipoSH function is defined as
\ba
\{Y_{\ell_1} (\hat{k}) \! \otimes \! Y_{\ell_2} (\hat{n})\}_{LM} &=&  \sum_{m_1 m_2} \mathcal{C}^{L M}_{\ell_1 m_1 \ell_2 m_2}  \nonumber \\
&\times& Y_{\ell_1 m_1} (\hat{k}) \; Y_{\ell_2 m_2} (\hat{n}). \nonumber
\ea

\noindent The pre-factors in equation (\ref{deltadef}) are the normalization terms associated with the CMB brightness fluctuation.
The deviations from statistical isotropy is associated with non-zero values of $L,\; L>0$. To check for the statistical
isotropy limit i.e. $L=0$ \& $M=0$, we use equation (8.5.1) from \cite{varsha} 
\be
\mathcal{C}^{0 0}_{\ell_1 m_1 \ell_2 m_2} = (-1)^{\ell_1 - m_1} \frac{\delta_{\ell_1 \ell_2} \delta_{m_1 \,-m_2}}{\sqrt{2 \ell_1 +1}} ,
\ee
to recover equation (\ref{delta}) in section (\ref{sec:delta}).
\subsection{Angular Correlations}

Starting with equation (\ref{obsanis}) and the Fourier Transform relation from equation (\ref{fourier}), the SH coefficients $a_{\ell m}$ 
for the general case are
\ba
a_{\ell m} &=& 4 \pi \int \frac{d^3 k}{(2\pi)^3} \; \phi(\vec{k}) \sum_{\ell_1 m_1 L M}\beta_{\ell_1 \ell} \;
\tilde{\Delta}_{\ell_1 \ell} ^{L M}(k,\tau) \nonumber \\
& \times& \mathcal{C}^{L M}_{\ell_1 m_1 \ell m} \; Y_{\ell_1 m_1}(\hat{k})
\label{alm}
.\ea
The angular correlations turn out to be
\ba
\langle a_{\ell m} a_{\ell' m'}^*\rangle &=& (4 \pi)^2 \int \int \frac{d^3 k}{(2\pi)^3} \frac{d^3 k'}{(2\pi)^3} 
\langle \phi(\vec{k})\phi^*(\vec{k'})\rangle \nonumber \\
&\times &\sum_{\ell_1 m_1 L M} \sum_{\ell_2 m_2 L' M'} \beta_{\ell_1 \ell}  \beta_{\ell_2 \ell'}^* \nonumber \\
&\times &\mathcal{C}^{L M}_{\ell_1 m_1 \ell m} \mathcal{C}^{L' M'}_{\ell_2 m_2 \ell' m'} \;
Y_{\ell_1 m_1}(\hat{k}) Y^*_{\ell_2 m_2}(\hat{k'}) \nonumber \\
&\times& \tilde{\Delta}_{\ell_1 \ell} ^{L M}(k,\tau) [\tilde{\Delta}_{\ell_2 \ell'} ^{L' M'}(k',\tau)]^*
\label{angcorr}
.\ea
The most general power spectrum (under statistical homogeneity) depends on the direction $\hat{k}$,
\be
\langle \phi(\vec{k})\phi^*(\vec{k'})\rangle = P(\vec{k}) \delta(\vec{k} - \vec{k'}) .
\ee
Further, it is useful to parametrize the directional dependence of $\hat{k}$ in $P(\vec{k})$ as \cite{pullen}
\be
P(\vec{k}) =  P_0(k) \left[1 + \sum_{\mathfrak{l} > 0} \sum_{\mathfrak{m}=-\mathfrak{l}}^{\mathfrak{l}} g_{\mathfrak{l} \mathfrak{m}}(k) \; 
Y_{\mathfrak{l} \mathfrak{m}}(\hat{k}) \right] ,
\label{k-ps}
\ee
where the first term with $\mathfrak{l} =0$ represents the statistical homogeneous and isotropic primordial power spectrum. 

\noindent For a directional dependent power spectrum, the angular correlations of temperature anisotropy can be written as
\ba
\langle a_{\ell m} a_{\ell' m'}^*\rangle & = & 4 \pi \!\! \int \frac{k^2 dk}{2\pi^2} P_0(k) \!\!\!\!\!\! \sum_{\ell_1 m_1 L M L' M'} \left[ D_1 
\tilde{\Delta}_{\ell_1 \ell} ^{L M}(k,\tau) \right . \nonumber \\
& \times & [\tilde{\Delta}_{\ell_1 \ell'} ^{L' M'}(k,\tau)]^* + \sum_{\ell_2 m_2 \mathfrak{l} \mathfrak{m}} D_2 \; g_{\mathfrak{l} \mathfrak{m}}(k) \nonumber \\
& \times & \left. \tilde{\Delta}_{\ell_1 \ell} ^{L M}(k,\tau) \; [\tilde{\Delta}_{\ell_2 \ell'}^{L' M'}(k,\tau)]^* \; 
\mathcal{Y}^{\ell_2 m_2}_{\ell_1 m_1, \mathfrak{l} \mathfrak{m}} \right ] 
\label{k-dep},
\ea
where
\ba
D_1 &=& \beta_{\ell_1 \ell} \; \beta^*_{\ell_1 \ell'}  \; \cg L M {\ell_1} {m_1} {\ell} m  \; \cg {L'} {M'} {\ell_1} {m_1} {\ell'} {m'} , \nonumber \\
D_2 &=& \beta_{\ell_1 \ell} \;  \beta^*_{\ell_2 \ell'}  \;  \cg L M {\ell_1} {m_1} {\ell} m  \;  \cg {L'} {M'} {\ell_2} {m_2} {\ell'} {m'}, \nonumber \\
\mathcal{Y}^{\ell_2 m_2}_{\ell_1 m_1 , \mathfrak{l} \mathfrak{m}} &=& \int d\Omega_{\hat{k}} \;  Y_{\ell_1 m_1}(k) \; Y^*_{\ell_2 m_2}(k) \; 
Y_{\mathfrak{l} \mathfrak{m}}(k) \nonumber \\
&=& \frac{\Pi_{\ell_1 \mathfrak{l}}}{\sqrt{4\pi} \; \Pi_{\ell_2}} \cg {\ell_2} {0} {\ell_1} {0} {\mathfrak{l}} {0} 
\cg {\ell_2} {m_2} {\ell_1} {m_1} {\mathfrak{l}} {\mathfrak{m}} \nonumber ,
\ea
where we have used the expression for integral of three spherical harmonics as in  equation (5.9.4) from \cite{varsha}. 

\noindent Here $\Pi_{\ell_1\ell_2 ...\ell_n} = [(2\ell_1 +1)(2\ell_2 +1)...(2\ell_n +1)]^{\frac{1}{2}}$ has been defined for convenient notational simplicity.

\noindent The case $L=0$ reduces equation (\ref{k-dep}) to that in the analysis \cite{pullen} 
\ba
\langle a_{\ell m} a_{\ell' m'}^*\rangle  &=&   \frac{(-1)^{\ell}}{\Pi_{\ell}} A^{00}_{\ell \ell'} \delta_{\ell \ell'} \delta_{m m'} \nonumber \\  
&+& (-1)^{\ell'+m'}  \sum_{\mathfrak{l},\mathfrak{m}}
A^{\mathfrak{l} \mathfrak{m}}_{\ell \ell'} \;\;  \cg {\mathfrak{l}} {\mathfrak{-m}}  {\ell} {m} {\ell'} {m'} ,
\label{gencov}
\ea
where statistical anisotropy is quantified by the BipoSH coefficients \cite{hajian, hajian-2004,hajian-2006, souradeep}, defined as 
a tensor product of the spherical harmonic coefficients $a_{\ell m}$ and $a_{\ell'm'}$,
\ba
A^{JN}_{\ell \ell'} &=& \sum_{m m'} \langle a_{\ell m} a_{\ell'm'}^* \rangle (-1)^{m'} \cg J N {\ell} {m} {\ell'} {-m'} \nonumber \\
&=& \left\{a_{\ell} \otimes a_{\ell'} \right\}_{JN} \; \nonumber \\
A_{\ell \ell}^{00} &=& (-1)^{\ell} \; \Pi_{\ell} \; C_{\ell}
\label{biposh}
\ea
Here, $C_{\ell}$ is the usual CMB power spectrum for the SI case. Directional dependent $P(\hat{k})$ \cite{pullen},
introduces the second term in equation (\ref{gencov})
\ba
A^{\mathfrak{l} \mathfrak{m}}_{\ell \ell'} &=&  \sqrt{4\pi} (-i)^{\ell+\ell'} \frac{\Pi_{\ell}\Pi_{\ell'}}{\Pi_{\mathfrak{l}}}\int \frac{k^2 dk}{2\pi^2} \; 
P_0(k) g_{\mathfrak{l} \mathfrak{m}} \nonumber \\
&\times& \cg {\mathfrak{l}} {0} {\ell} 0 {\ell'} {0} \left\{\tilde{\Delta}^0_{\ell \ell} \otimes \tilde{\Delta}^{0}_{\ell' \ell'}\right\}_{00} , 
\label{si-biposh}
\ea
where a corresponding tensor product in Bipolar harmonic space for the indices $L$ and $L'$ of CMB brightness fluctuations is defined as
\ba
\left\{\tilde{\Delta}^L_{\ell_1 \ell} \otimes \tilde{\Delta}^{L'}_{\ell_2 \ell'}\right\}_{J N} &=& \sum_{M M'} (-1)^{M'}  
\cg {J} {N} {L} {M} {L'} {-M'} \nonumber \\
& \times & \tilde{\Delta}_{\ell_1 \ell} ^{L M}(k,\tau) [\tilde{\Delta}_{\ell_2 \ell'} ^{L' M'}(k,\tau)]^* \! .
\label{new-biposh}
\ea

\vspace{2mm}

\noindent In general for SI violations $(L>0)$, the BipoSH coefficients can be expressed using the angular correlations in equation (\ref{k-dep}) 
as shown in Appendix \ref{app0} as
\ba
A^{J N}_{\ell \ell'} & = &  \left(A^{J N}_{\ell \ell'}\right)_{\mathfrak{l}=0} + \sqrt{4 \pi} \int \frac{k^2 dk}{2 \pi^2} P_0(k)
\nonumber \\
& \times & \sum_{\mathfrak{l} \ell_1 \ell_2 L L'} \!\!\! \Pi_{\mathfrak{l} \ell_1  L L'} (-1)^{\mathfrak{l} + L'+ \ell' + \ell_1 - \ell_2} \beta_{\ell_1 \ell} 
\beta_{\ell_2 \ell'}^* \nonumber \\
& \times &  \cg {\ell_2} {0} {\ell_1} {0} {\mathfrak{l}} {0}  \sum_{\mathfrak{m}} g_{\mathfrak{l} \mathfrak{m}} \sum_{\ell_3 m_3} 
\wninej {J} {\ell_3} {\mathfrak{l}} {\ell} {L} {\ell_1} {\ell'}  {L'} {\ell_2} \nonumber \\
& \times &  \cg {J} {N} {\ell_3} {m_3} {\mathfrak{l}} {\mathfrak{m}}   
\left\{\tilde{\Delta}^L_{\ell_1 \ell} \otimes \tilde{\Delta}^{L'}_{\ell_2 \ell'}\right\}_{\ell_3 m_3}. 
\label{gen-biposh}
\ea
The first term on the right-hand side of equation (\ref{gen-biposh}), is the contribution due to statistically isotropic primordial power spectrum 
i.e. $\mathfrak{l} = 0$ terms in equation (\ref{k-ps}). The second term gives the contribution to the Bipolar coefficients due to $\mathfrak{l} > 0$ in equation (\ref{k-ps}). 
The term in the first braces is the Wigner-9j symbol \cite{varsha} which is related to the coefficients of transformations between different coupling 
schemes of four angular momenta and satisfies the triangular conditions for the triads $(J \ell_3 \mathfrak{l}),
(\ell L \ell_1), (\ell'L'\ell_2), (J\ell\ell'), (\ell_3 L L')$ and $(\mathfrak{l} \ell_1 \ell_2)$. 

\noindent The SI case $L=0$ reduces the BipoSH coefficients in equation (\ref{gen-biposh}) to equation (\ref{si-biposh}). 

\vspace{2mm}

\noindent To evaluate the first term $\left(A^{J N}_{\ell \ell'}\right)_{\mathfrak{l}=0}$ in equation (\ref{gen-biposh}), we consider
statistically isotropic primordial perturbations $\langle \phi(\vec{k})\phi^*(\vec{k'})\rangle = P_0(k) \delta(\vec{k} - \vec{k'})$
and express the angular correlations  as
\ba
\langle a_{\ell m} a_{\ell' m'}^*\rangle &=& 4 \pi \int \frac{k^2 dk}{2 \pi^2} P_0(k) \sum_{\ell_1 m_1 L M L' M'}
\beta_{\ell_1 \ell} \: \beta_{\ell_1 \ell'}^* \:  \nonumber \\
&\times & \tilde{\Delta}_{\ell_1 \ell} ^{L M}(k,\tau)
\: [\tilde{\Delta}_{\ell_1 \ell'} ^{L' M'}(k,\tau)]^* \nonumber \\
&\times& \cg L M {\ell_1} {m_1} {\ell} m \: \cg {L'} {M'} {\ell_1} {m_1} {\ell'} {m'}
\label{si-corr}
.\ea
To check for the statistical isotropic case, we put $L=0$ \& $M=0$ and recover equation (\ref{corr}) in section (\ref{sec:corr}).

\noindent As shown in detail in Appendix \ref{app0}, the BipoSH coefficients in equation (\ref{biposh}), can be expressed using the angular correlations in 
equation (\ref{si-corr}) for statistically isotropic primordial perturbations as 
\ba
(A^{J N}_{\ell \ell'})_{\mathfrak{l}=0} &=& 4 \pi \int \frac{k^2 dk}{2 \pi^2} P_0(k) \sum_{\ell_1 L L'} (-1)^{\ell_1+L+L'+J}  \nonumber \\
& \times & \Pi_{L L'} \:\: \beta_{\ell_1 \ell} \:\: \beta_{\ell_1 \ell'}^* \: \wsixj {L} {J} {L'} {\ell'} {\ell_1} {\ell} \nonumber \\
& \times & \left\{\tilde{\Delta}^{L}_{\ell_1 \ell} \otimes \tilde{\Delta}^{L'}_{\ell_1 \ell'}\right\}_{J N}.
\label{re-biposh}
\ea
The first term in braces is the Wigner-6j symbol which is related to the coefficients of transformations between different coupling 
schemes of three angular momenta. These vanish unless the triangular conditions \cite{varsha} are fulfilled for the triads 
$(LJL'),(L'\ell'\ell_1),(J\ell'\ell)$ and $(\ell L \ell_1)$.

\noindent We consider low Bipolar deviations from SI i.e. $L,L'$ and $J << \ell,\ell',\ell_1$  in equation (\ref{re-biposh}) and use the asymptotic relation for 
Wigner-6j functions given by equation (9.9.1) from \cite{varsha}. We find that the asymptotic limit to these BipoSH coefficients are
\ba
(A^{J N}_{\ell \ell'})_{\mathfrak{l}=0} & \approx & 4 \pi \int \frac{k^2 dk}{2 \pi^2} P_0(k) \sum_{\ell_1 L L'} (-1)^{\ell + \ell'+ \ell_1+L+L'}  \nonumber \\
&\times & \frac{\Pi_{L L'}}{\sqrt{2\ell_1} \;  \Pi_{J}} \:\: \beta_{\ell_1 \ell} \:\: \beta_{\ell_1 \ell'}^* \:\: 
\cg {J} {(\ell-\ell')} {L} {(\ell-\ell_1)} {L'} {(\ell_1-\ell')} \nonumber \\
&\times &  \left\{\tilde{\Delta}^L_{\ell_1 \ell} \otimes \tilde{\Delta}^{L'}_{\ell_1 \ell'}\right\}_{J N} \; .
\label{asymp-biposh}
\ea
For diagonal brightness fluctuations i.e $\ell_1 = \ell$ and $\ell_1 = \ell'$, the BipoSH coefficients themselves turn out to be diagonal in multipole space 
\ba
(A^{J N}_{\ell \ell})_{\mathfrak{l}=0} & \approx & 4 \pi \int \frac{k^2 dk}{2 \pi^2} P_0(k) \sum_{L L'} (-1)^{L+L'+1} \;
\frac{\Pi_{\ell \ell L L'}}{\Pi_{J}} \nonumber \\
& \times & \cg {J} {0} {L} {0} {L'} {0} \left\{\tilde{\Delta}^L_{\ell \ell} \otimes \tilde{\Delta}^{L'}_{\ell \ell}\right\}_{J N} . 
\label{diag-asymp-biposh}
\ea

\section{Evolution in the free-streaming regime}
\subsection{Generalized evolution equation}
We find the moments of the CMB brightness fluctuation to be
\ba
\tilde{\Delta}_{\ell_1 \ell_2} ^{L M}(k,\tau) &=& \frac{1}{4 \pi} \frac{1}{\beta_{\ell_1 \ell_2}} \int \int d\Omega_{\hat{n}} \; d\Omega_{\hat{k}}
\; \tilde{\Delta}(\vec{k}, \hat{n}, \tau) \nonumber \\
&\times & \{Y_{\ell_1} (\hat{k}) \otimes Y_{\ell_2}(\hat{n})\}_{LM},
\label{del-moment}
\ea
starting with equation (\ref{deltadef}) and the orthonormality condition of BipoSH,
\ba
\int \!\!\! \int d\Omega_{\hat{n}} d\Omega_{\hat{k}} &&\!\!\!\!\!\!\!\!\!\!\! \{Y_{\ell_1} (\hat{k}) \otimes Y_{\ell_2}(\hat{n})\}_{LM}
\{Y_{\ell_3} (\hat{k}) \otimes Y_{\ell_4} (\hat{n})\}_{L'M'}^* \nonumber \\
&=& \delta_{\ell_1 \ell_3} \;\delta_{\ell_2 \ell_4} \:\delta_{L L'}
\delta_{M M'}.
\ea
In the free streaming regime, using the plane wave approximation as in equations (\ref{free}) and (\ref{plane}), the most
general evolution equation turns out to be
\ba
\tilde{\Delta}_{\ell_1 \ell_2} ^{L M}(k,\tau) &=& \frac{1}{\beta_{\ell_1 \ell_2}} \sum_{\ell \ell_3 \ell_4 L' M'} (-i)^\ell \;
\Pi_{\ell\ell} \; j_\ell(k \Delta \tau) \; \beta_{\ell_3 \ell_4} \nonumber  \\
&\times &\sum_{m_1 m_2 m_3 m_4} \!\!\!\!\!\!\! \tilde{\Delta}_{\ell_3 \ell_4} ^{L M}(k,\tau_s) \: \mathcal{C}^{L M}_{\ell_1 m_1 \ell_2 m_2} \:
\mathcal{C}^{L' M'}_{\ell_3 m_3 \ell_4 m_4} \nonumber  \\
& \times &\int \!\!\! \int d\Omega_{\hat{n}}\; d{\hat{k}} \; P_\ell(\hat{k} \cdot \hat{n})\: Y_{\ell_1 m_1}^*({\hat{k}}) \:
Y_{\ell_2 m_2}^*({\hat{n}}) \nonumber \\
& \times& Y_{\ell_3 m_3}({\hat{k}}) \: Y_{\ell_4 m_4}({\hat{n}})
\label{evolgen}
.\ea
As shown in detail in Appendix \ref{appA}, this can be further simplified to
\ba
\tilde{\Delta}_{\ell_1 \ell_2} ^{L M}(k,\tau)&=&\sum_{\ell \ell_3 \ell_4} (-i)^{\ell} \; \Pi_{\ell \ell \ell_1 \ell_4}
\frac{\beta_{\ell_3 \ell_4}}{\beta_{\ell_1 \ell_2}} \; (-1)^{\ell_3 + \ell_4 + L}  \nonumber \\
&\times & j_\ell(k \Delta \tau) \tilde{\Delta}_{\ell_3 \ell_4} ^{L M}(k,\tau_s) \; \cg {\ell_3} 0 {\ell} 0 {\ell_1} 0 \;
\cg {\ell_2} 0 {\ell} 0 {\ell_4} 0  \nonumber \\
&\times&\wsixj {\ell_1} L {\ell_2} {\ell_4} {\ell} {\ell_3} \nonumber \\
\ea
The generalized evolution equation thus can be expressed so as to structurally resemble the evolution equation for the SI case 
in equation (\ref{evolve}).
\ba
\tilde{\Delta}_{\ell_1 \ell_2} ^{L M}(k,\tau) &=& \sum_{\ell \ell_3 \ell_4}  C(L,\ell,\ell_1,\ell_2) \;
j_\ell(k \Delta \tau) \nonumber \\
&\times& \tilde{\Delta}_{\ell_3 \ell_4} ^{L M}(k,\tau_s) ,
\label{simevol}
\ea
where
\ba
C(L,\ell,\ell_1,\ell_2) &=& (-i)^{\ell} \; \Pi_{\ell \ell \ell_1 \ell_4}
\frac{\beta_{\ell_3 \ell_4}}{\beta_{\ell_1 \ell_2}} (-1)^{\ell_3 + \ell_4 + L} \nonumber \\
&\times&  \cg {\ell_3} 0 {\ell} 0 {\ell_1} 0 \;
\cg {\ell_2} 0 {\ell} 0 {\ell_4} 0 \; \wsixj {\ell_1} L {\ell_2} {\ell_4} {\ell} {\ell_3}.
\label{c-plot}
\ea
\begin{figure}[htb]
\centering
\subfigure[\;Case 1: Plot of $C(L,\ell,\ell_1,\ell_2)$ for $\tilde{\Delta}_{11} ^{L M}(k,\tau_s)$; with initial SH multipoles $\;\ell_3 = \ell_4 = 1$, 
the final SH multipoles being $\ell = \ell_1 - 1 (= \ell_2 - 1)$ and the index of deviation from statistical isotropy 
$L = 0$ (blue, solid), $L = 1$ (red, dashed) and $L = 2$ (black, dot-dashed) respectively. ]
{
    \label{fig1}
    \includegraphics[width=8cm]{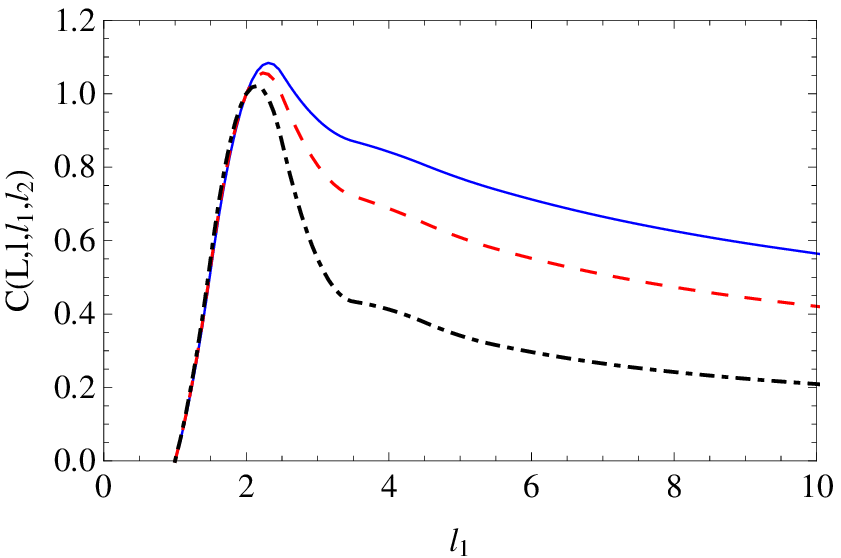}
}
\hspace{0.5cm}
\vspace{5mm}
\subfigure[\;Case 2: Plot of $C(L,\ell,\ell_1,\ell_2)$ for $\tilde{\Delta}_{22} ^{L M}(k,\tau_s)$; with initial multipoles $\;\ell_3 = \ell_4 = 2$, 
the final multipoles being $\ell = \ell_1 (= \ell_2)$ and the index of deviation from statistical isotropy $L = 0$ (blue, solid), 
$L = 1$ (red, dashed) and $L = 2$ (black, dot-dashed) respectively. ]
{
    \label{fig2}
    \includegraphics[width=8cm]{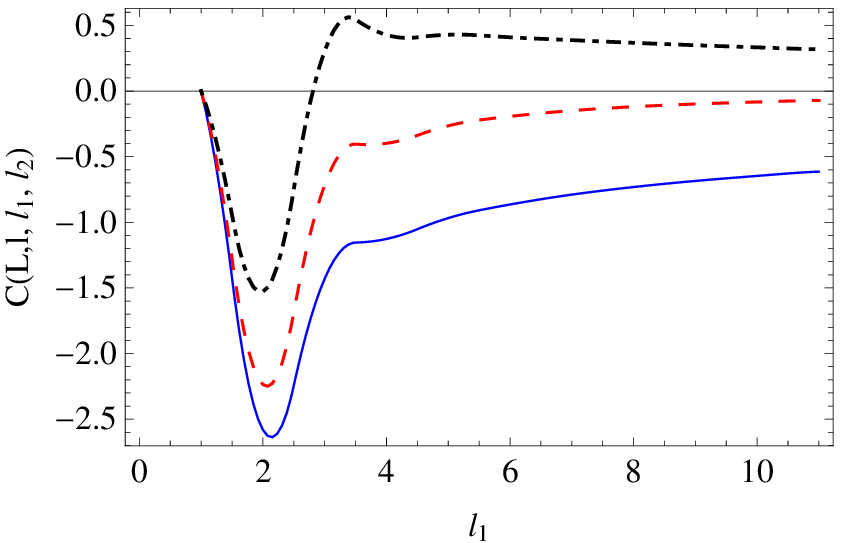}
}
\caption{Evolution of the coefficient of the spherical Bessel functions, $C(L,\ell,\ell_1,\ell_2)$ with multipole moment $\ell_1$ as in equation 
(\ref{simevol}), for non-zero diagonal terms of a unit normalized $\tilde{\Delta}_{\ell_3 \ell_4} ^{L M}(k,\tau_s)$ }
\label{fig:a}
\end{figure}

\begin{figure}[ht]
\centering
\subfigure[\;Case 1: Plot of $C(L,\ell,\ell_1,\ell_2)$ for $\tilde{\Delta}_{10} ^{L M}(k,\tau_s)$; with initial multipoles 
$\;\ell_3 = 1, \; \ell_4 = 0$, the final multipoles being $\ell = \ell_2 = \ell_1 + 1 $\;(blue, solid) and  $\ell = \ell_2 = \ell_1 - 1 $\;(red, dashed) respectively. 
The terms are non-zero only when the index of deviation from statistical isotropy $L = |\ell_3 - \ell_4| = 1$. ]
{
    \label{fig3}
    \includegraphics[width=8cm]{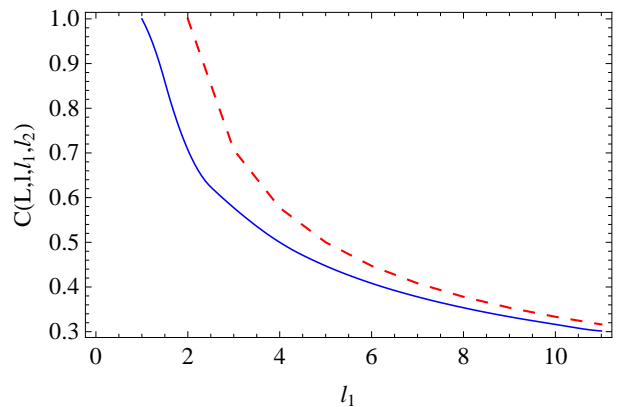}
}
\hspace{0.5cm}
\vspace{5mm}
\subfigure[\;Case 2: Plot of $C(L,\ell,\ell_1,\ell_2)$ for $\tilde{\Delta}_{20} ^{L M}(k,\tau_s)$; with initial multipoles 
$\;\ell_3 = 2,\; \ell_4 = 0$, the final multipoles being $\ell = \ell_2 = \ell_1  $\;(blue, solid) and  $\ell = \ell_2 = \ell_1 - 2 $\;(red, dashed) respectively.
The terms are non-zero only when the index of deviation from statistical isotropy $L = |\ell_3 - \ell_4| = 2$.]
{
    \label{fig4}
    \includegraphics[width=8cm]{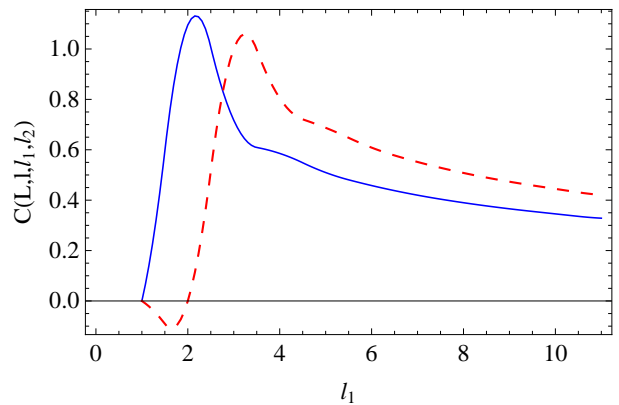}
}
\caption{Evolution of the coefficient of the spherical Bessel functions, $C(L,\ell,\ell_1,\ell_2)$ with multipole moment $\ell_1$ as in equation 
(\ref{simevol}), for non-zero off-diagonal terms of a unit normalized $\tilde{\Delta}_{\ell_3 \ell_4} ^{L M}(k,\tau_s)$ }
\label{fig:b}
\end{figure}

\noindent Setting $L=0, M=0$ we recover the statistical isotropic case as in equation (\ref{evolve}), section (\ref{sec:evol}).

\noindent The transfer of power of the statistical anisotropic terms to higher SH multipoles $\ell_1$ \& $\ell_2$ due to free streaming is illustrated in 
figures (\ref{fig:a}) and (\ref{fig:b}). Starting from a unit normalized $\tilde{\Delta}_{\ell_3,\ell_4} ^{L M}(k,\tau_s)$, we plot the evolution of the 
coefficients in equation (\ref{c-plot}),  $C(L,\ell,\ell_1,\ell_2)$ with $\ell_1$ for specific values of the SH multipole moments $\ell_3$ and $\ell_4$. 
We find that the values of $\ell$ and $\ell_2$ are constrained by the values of $\ell_1$ due to the triangular inequalities of the Clebsch-Gordon coefficients. 

\noindent Figure \ref{fig:a} shows the evolution of these coefficients for diagonal, unit normalized CMB brightness fluctuations namely 
$\tilde{\Delta}_{11} ^{L M}(k,\tau_s)$ and $\tilde{\Delta}_{22} ^{L M}(k,\tau_s)$ with $L = 0,1,2$, for possible values of $\ell$ and $\ell_2$. 

\noindent Figure \ref{fig:b} shows the evolution of the coefficients for off-diagonal, unit normalized $\tilde{\Delta}_{10} ^{L M}(k,\tau_s)$ and 
$\tilde{\Delta}_{20} ^{L M}(k,\tau_s)$, for possible values of $L, \ell$ and $\ell_2$. The coefficients for off-diagonal
$\tilde{\Delta}_{\ell_3,\ell_4} ^{L M}(k,\tau_s)$ vanish for $L=0$ i.e. the statistical isotropic case. 

\subsection{SI violation at large multipoles}

The ability to measure violation of statistical isotropy at low SH
multipoles is largely compromised by cosmic variance. At larger SH
multipoles, the effect due to violation of SI would become prominent.
The previous section shows that power is transferred from small to
large SH multipoles during free-streaming. This opens the door to more
readily measurable SI violations arising from SI violation induced
due to physical processes (e.g., presence of magnetic fields,or, other
breakdown of rotational symmetries) in the baryon-photon plasma.

Due to the tight coupling in the baryon-photon plasma prior to
$\tau_s$, primordial SI violation can be expected to be limited to
small SH multipoles.  It is illuminating then to obtain an expression
for the free-streaming of BipoSH brightness fluctuations at small SH
multipole moments at the last scattering to large SH multipoles at the
present epoch for which we essentially evaluate the asymptotic limit
of the CMB brightness fluctuations today.

\noindent In equation (\ref{simevol}) we take $\ell_1$ \& $\ell_2$ to be large compared to $\ell_3$ \&  $\ell_4$. From symmetry of Wigner-6j symbols
\be
\wsixj {\ell_1} {\ell}  {\ell_3} {\ell_4} L {\ell_2} = \wsixj {\ell_4} L {\ell_3} {\ell_1} {\ell} {\ell_2}.
\ee
We evaluate the asymptotic limit with arbitrary values of $\ell$, $\ell_1$, $\ell_2$, $\ell_3$ and $\ell_4$ using equation (9.9.1) from \cite{varsha} as, 
\ba
\wsixj {\ell_4} L {\ell_3} {R+d} {R+e} {R+f} &\approx& \frac{(-1)^{\ell_4+L+d+e}}{\sqrt{2\ell} \:\Pi_{\ell_3}} \nonumber \\
&\times& \cg {\ell_3} {(d-e)} {\ell_4} {(f-e)} L {(d-f)}
\label{asymp}
\ea
where $R$ is large and
\ba
d &=& \ell_1 -R \nonumber \\
e &=& \ell -R  \nonumber \\
f &=& \ell_2 -R  . \nonumber
\ea
Thus for $\ell_1, \ell_2 >>  \ell_3, \ell_4$, equation (\ref{asymp}) becomes
\be
\wsixj {\ell_4} L {\ell_3} {\ell_1} {\ell} {\ell_2} \approx \frac{(-1)^{\ell_4+L+\ell_1 + \ell -2R}}{\sqrt{2\ell} \:\Pi_{\ell_3}}
\cg {\ell_3} {(\ell_1-\ell)} {\ell_4} {(\ell_2-\ell)} L {(\ell_1-\ell_2)}
\label{sixj}
.\ee
Putting the expression for the Wigner-6j symbols from equation (\ref{sixj}), the asymptotic limit of the generalized evolution equation (\ref{simevol})
takes the form
\ba
\tilde{\Delta}_{\ell_1 \ell_2} ^{L M}(k,\tau)&=&\sum_{\ell \ell_3 \ell_4} (-i)^{\ell} \; \frac{\Pi_{\ell \ell \ell_1 \ell_4}}{\sqrt{2 \ell} \Pi_{\ell_3}}
\frac{\beta_{\ell_3 \ell_4}}{\beta_{\ell_1 \ell_2}} \; (-1)^{\ell + \ell_1 + \ell_3}  \nonumber \\
&\times & j_\ell(k \Delta \tau) \tilde{\Delta}_{\ell_3 \ell_4} ^{L M}(k,\tau_s) \; \cg {\ell_3} 0 {\ell} 0 {\ell_1} 0 \;
\cg {\ell_2} 0 {\ell} 0 {\ell_4} 0  \nonumber \\
&\times& \cg {\ell_3} {(\ell_1-\ell)} {\ell_4} {(\ell_2-\ell)} L {(\ell_1-\ell_2)} \, .
\label{largel}
\ea 
Equation (\ref{largel}) depicts how power in SI violating terms at
small SH multipoles $\ell_3, \ell_4$ at the last scattering,
free stream to higher SH multipoles $\ell_1, \ell_2$ at the present
epoch. Note the structural similarity to equation (\ref{evolve}) for
statistical isotropic case. 

\vspace{2mm}

\noindent It is useful to provide explicit expressions for equation (\ref{largel}) in two particular cases, when
the asymptotic moments of the CMB brightness fluctuation contains only diagonal terms and off-diagonal terms respectively.

\noindent The evolution equation which involves only the diagonal terms of the moments of the brightness fluctuation at last scattering are
\ba
\tilde{\Delta}_{\ell_1, \ell_1} ^{L M}(k,\tau) &=& \sum_{\ell} C_1(L, \ell,\ell_1) \; j_\ell(k \Delta \tau) \nonumber \\
&\times& \tilde{\Delta}_{\ell_1-\ell ,\; \ell_1-\ell} ^{L M}(k,\tau_s) \;\; ,
\ea
with
\ba
C_1(L,\ell, \ell_1) &=&  \frac{\Pi_{\ell \ell \ell_1 \ell_3}}{\sqrt{2\ell}}  
\left[ \frac{\ell_1!}{\ell! \ell_3!} \right] ^2 
\frac{(2\ell)!(2\ell_3+1)!}{(2\ell_1+1)!} \; \delta_{\ell_3,\ell_1-\ell} \nonumber \\
&\times& \sqrt{\frac{(2\ell_3)!}{(2\ell_3 -L)!} \frac{(2\ell_3)!}{(2\ell_3+L+1)!}} 
\label{diag}
\ea
where $\ell,\ell_1 >> |\ell_1 -\ell|$. The details are given in Appendix \ref{appB}. The term under the square-root captures the 
$L$ dependence of the free-streaming of $L>0$ terms.

\vspace{2mm}

\noindent The evolution equation which involves only the off-diagonal terms of the moments of the brightness fluctuation at last scattering are
\ba
\tilde{\Delta}_{\ell_1, \ell_2} ^{L M}(k,\tau) &=& \sum_{\ell} C_2(\ell, \ell_1, \ell_2) \: j_{\ell}(k\Delta \tau) \nonumber \\
& \times & \tilde{\Delta}_{\ell_1 -\ell, \ell_2-\ell} ^{L M}(k,\tau_s) \delta_{L, \ell_1 - \ell_2}
,\ea
with
\ba
C_2(\ell, \ell_1, \ell_2) &=&  \frac{(-1)^{\frac{\ell_1 + \ell_2}{2}}}{\sqrt{2\ell}} 
\frac{\Pi_{\ell \ell \ell_1 \ell_2}}{\Pi_{\ell_1-\ell}} \: \sqrt{\frac{\ell_1+\ell_2-2\ell+1}{\ell_1+\ell_2+1}} \nonumber \\
& \times & \frac{(2\ell)! \ell_1 ! \ell_2 ! }{(\ell!)^2 (\ell_1 - \ell)! (\ell_2 - \ell)!} \nonumber \\
& \times & \left[\frac{(2\ell_1 -2 \ell + 1)! (2\ell_2 -2 \ell + 1)!}{(2\ell_1+1)! (2\ell_2+1)!} \right] ^{\frac{1}{2}}
\label{offdiag}
.\ea
where $\ell,\; \ell_n >> |\ell_n -\ell|,\; L$ with $n = 1,2$. The details are given in Appendix \ref{appC}. 
As in equation (\ref{diag}), the term under the square-root captures the $L$ dependence of the free-streaming of $L>0$ terms.

\subsection{SI violating physical effects at last scattering} 

The patterns of the CMB temperature field i.e. the angular correlations observed today are traced back to inhomogeneities at the last scattering surface.
In the tight coupling regime of the baryon-photon fluid, one expects power only at small SH multipoles of the $\tilde{\Delta}_{\ell_1 \ell_2} ^{L M}(k,\tau_s)$.
The generalized evolution equation of the CMB brightness fluctuations (\ref{evolgen}) 
free-streams this power at small SH multipoles in both SI and non-SI moments to corresponding SI and non-SI moments with same Bipolar moment $L$. 
Any observed violation of SI today is easier to interpret as generalized moments arising due to simple physics just beyond the fluid approximation regime. 
In this section, we illustrate this point explicitly for SI violation in the CMB anisotropy in the presence of a homogeneous magnetic field at last scattering. 

\noindent The SH coefficients $a_{\ell m}$ can be expressed in terms of the CMB brightness fluctuations at last scattering using equation (\ref{alm})
and the generalized evolution equation (\ref{simevol}) as
\ba
a_{\ell m} &=& 4 \pi \int \frac{d^3 k}{(2\pi)^3} \; \phi(\vec{k}) \sum_{\ell_1 m_1 L M} \beta_{\ell_1 \ell} \;
\mathcal{C}^{L M}_{\ell_1 m_1 \ell m} \; Y_{\ell_1 m_1}(\hat{k}) \nonumber \\
&\times&  \sum_{\ell_2 \ell_3 \ell_4}  C(L,\ell_2,\ell_1,\ell) \; j_{\ell_2}(k \Delta \tau) \; \tilde{\Delta}_{\ell_3 \ell_4} ^{L M}(k,\tau_s),
\label{alm-ls}
\ea
where $C(L,\ell_2,\ell_1,\ell)$ is defined in equation (\ref{c-plot}). 

\noindent Hence, in general $\langle a_{\ell m} a_{\ell' m'} \rangle$ correlations measured at present are related to 
$\langle \tilde{\Delta}_{\ell_1 \ell_2} ^{L M}(k,\tau_s) \tilde{\Delta}_{\ell_3 \ell_4}^{L' M'}(k,\tau_s) \rangle$
correlation between the generalized Boltzmann fluctuations at the last scattering surface. 

\noindent In particular, SI violation encoded in the off-diagonal correlation $\langle a_{\ell m} a_{\ell' m'} \rangle$ 
(and non-zero BipoSH $A^{LM}_{\ell \ell'}$, $L>0$) are related as
\ba
A^{JN}_{\ell \ell'} \sim \left\{\tilde{\Delta}^{L}_{\ell_1 \ell}(k,\tau_s) \otimes \tilde{\Delta}^{L'}_{\ell_2 \ell'}(k,\tau_s)\right\}_{JN} 
\ea
as in equations (\ref{re-biposh}, \ref{asymp-biposh}, \ref{diag-asymp-biposh}) or more generally for different Bipolar coefficients $J'$ and $N'$ as in 
equation (\ref{gen-biposh}) when the power spectrum is also anisotropic.

\noindent Using equation (\ref{deltadef}), correlations of the CMB brightness fluctuations are
\ba
\langle \Delta(\vec{k},\hat{n},\tau_s) \Delta(\vec{k},\hat{n'},\tau_s) \rangle \!\! &=& \!\! (4\pi)^2 \!\! \sum_{\ell_1 \ell_2 L M} \sum_{\ell_3 \ell_4 L' M'}
\beta_{\ell_1 \ell_2} \; \beta_{\ell_3 \ell_4}  \nonumber \\
&\times& \tilde{\Delta}^{LM}_{\ell_1 \ell_2}(k,\tau_s) \tilde{\Delta}^{L'M'}_{\ell_3 \ell_4}(k,\tau_s)  \nonumber \\
&\times&  \left\{ Y_{\ell_1}(\hat{k}) \otimes Y_{\ell_2}(\hat{n}) \right\}_{LM}  \nonumber \\
&\times& \left\{ Y_{\ell_3}(\hat{k}) \otimes Y_{\ell_4}(\hat{n'}) \right\}_{L'M'}.
\label{del-corr}
\ea

\vspace{2mm}

\noindent We illustrate the generality and power of our formalism using the case for a uniform magnetic field.
We show the correlations of the CMB brightness fluctuations in this particular case is sourced by the 
bipolar dipole ($L=1$) terms of equation (\ref{deltadef}) with $\ell_1 = \ell_2 =1$ where
\ba
\Delta(\vec{k},\hat{n},\tau_s) \! &=& 4\pi i \sqrt{3} \sum_M \tilde{\Delta}^{1M}_{11}(k,\tau_s) \left \{ Y_1(\hat{k}) \otimes Y_1(\hat{n}) \right \}_{1M} \nonumber \\
&=& 3i\sqrt{\frac{3}{2}} \sum_M \tilde{\Delta}^{1M}_{11}(k,\tau_s)\; \; (\hat{k} \boldsymbol{\times} \hat{n})_M ,
\label{magsi}
\ea
with $M=\{-1,0,+1\}$. Here $(\hat{k} \boldsymbol{\times} \hat{n})_M$ is the usual cross-product written as irreducible products of the rotation group \cite{varsha}.

\noindent Using standard vector identity \cite{varsha} and equation (\ref{free})
\ba
(\hat{n} \cdot \hat{n'})(\hat{k} \cdot \hat{k}) &-& (\hat{n} \cdot \hat{k})(\hat{n'} \cdot \hat{k}) \nonumber \\
&=& (\hat{n} \boldsymbol{\times} \hat{k}) \cdot (\hat{n'} \boldsymbol{\times} \hat{k}) ,
\ea
the temperature correlations in presence of a uniform magnetic field as discussed in \cite{durrer} (see equation A3) are given by 
\ba
\langle \Delta(\vec{k},\hat{n},\tau_s) \Delta(\vec{k},\hat{n'},\tau_s) \rangle &\propto & \tilde{\Delta}^{1M}_{11}(k,\tau_s)\; \;\tilde{\Delta}^{1M'}_{11}(k,\tau_s) \nonumber\\
&\times&  (\hat{n} \boldsymbol{\times} \hat{k}) \cdot (\hat{n'} \boldsymbol{\times} \hat{k}) .
\ea
Here the bipolar dipole terms of the CMB brightness fluctuation $\tilde{\Delta}^{1M}_{11}(k,\tau_s)\;\tilde{\Delta}^{1M'}_{11}(k,\tau_s)$ in equation (\ref{magsi}),
encapsulates the source term due to the presence of a uniform magnetic field \cite{durrer}.

\vspace{2mm}

\noindent In equation (\ref{alm-ls}), the bipolar dipole terms $\tilde{\Delta}_{11} ^{1 M}(k,\tau_s)$ gives rise to SH coefficients
\ba
a_{\ell m} &=& 4 \pi \int \frac{d^3 k}{(2\pi)^3} \; \phi(\vec{k}) \sum_{m_1  M} \beta_{1 \ell} \;
\mathcal{C}^{1 M}_{1 m_1 \ell m} \; Y_{1 m_1}(\hat{k}) \nonumber \\
&\times&  \sum_{\ell_3 } \left[(...)j_{\ell-1}(k \Delta \tau) + (...)j_{\ell+1}(k \Delta \tau)\right] \nonumber \\
&\times& \tilde{\Delta}_{1 1} ^{1 M}(k,\tau_s).
\ea
The angular correlations in this case turn out to be
\ba
\langle a_{\ell m} a^*_{\ell' m'} \rangle &=& (...)\; \delta_{\ell'\ell}  \delta_{mm'} + (...)\;\delta_{\ell'\:\ell \pm 2}  \delta_{mm'} .
\ea
It is interesting to note that the known diagonal ($\ell' = \ell$) and off-diagonal ($\ell' = \ell \pm 2$) correlations in presence of a homogeneous magnetic field
\cite{durrer,kahniashvili}, can be easily recovered in our approach. Work is in progress to relate other cases of SI violation 
originating in the physics at the last scattering surface using this formalism. 

\section{Conclusions}

The search for subtle statistical isotropy breakdown in the
universe is highly motivated by numerous theoretical scenarios. The
fluctuations in the cosmic microwave background is the arguably the
most promising observational probe of the SI of the universe.  The
violation of SI could have its origin not only in in anisotropic
primordial power spectrum, but also in the SI violation in the
fluctuations of the baryon-photon fluid at last scattering. SI
deviations generated by a general form of anisotropic primordial power
spectrum for isotropic Boltzmann functions has been studied in the
recent literature \cite{pullen}. \emph{This paper includes this equally important
possibility of a general SI breakdown in the CMB photon distribution
function.} We study the generalized case of SI violation in terms of
Bipolar spherical harmonic (BipoSH) brightness fluctuations,
substantially extending the scope of origin of SI violation solely
from the anisotropic primordial power spectrum.

The breakdown of SI in the CMB brightness fluctuation results in
off-diagonal terms in the SH space angular correlations $\langle a_{\ell
m} a_{\ell' m'} \rangle$, or, equivalently, in the coefficients the
Bipolar spherical harmonic (BipoSH) representation~\cite{hajian-2004,hajian-2006}.
We relate the measurable BipoSH coefficients to SI deviations in the
baryon-photon fluid, as well as, the primordial power spectrum.  The
observable BipoSH coefficients can be compactly expressed in terms
BipoSH brightness fluctuations terms through products of standard
Clebsch-Gordon coefficients and a Wigner-9j function. We also present
the expression for the simpler case of an isotropic primordial power
spectrum, where the BipoSH coefficients turn out to be given through a
compact combination of a Wigner-6j symbol and a Clebsch-Gordon
coefficient. We also provide the large SH multipole limit for these
coefficients for the terms encoding deviations from SI at low BipoSH
multipoles.

We obtain the generalized free-stream evolution equation for the SI
violation encoded in terms of the BipoSH brightness fluctuations
introduced in our work. We demonstrate that different modes BipoSH
brightness fluctuations at the present epoch have to evolve from same
Bipolar modes at the last scattering. The moments of the CMB
brightness fluctuations $\tilde{\Delta}^{LM}_{\ell_1 \ell_2}(\hat{k},
\tau_s)$ at last scattering are expected to non-zero at small values
of SH multipoles $\ell_1$ and $\ell_2$ due to tight coupling.
However, our results show that the power in these SI violating terms
at low SH multipoles would be transferred during free-stream evolution
to higher multipoles $\ell_3$ and $\ell_4$ in
$\tilde{\Delta}^{LM}_{\ell_3 \ell_4}(\hat{k}, \tau)$ at the present
epoch.  This is akin to the well known free-streaming evolution of
power in the SI brightness fluctuation at low SH multipole power at last
scattering to large SH multipole at present.  For clearly highlighting
the structural similarity, we present the evolution of BipoSH
brightness fluctuations in the asymptotic case of large values of the
final SH multipoles today relative to the initial SH multipoles at
last scattering. 

While many of the claimed observational evidence of SI breakdown,
such as the ``axis of evil'', ``north-south asymmetry'' etc., pertain
to relatively small values of the SH multipoles where the significance
is largely obscured by dominance of cosmic variance. However, SI violation 
at small SH multipoles in the baryon-photon plasma at last scattering would 
free-stream to large SH multipoles at present, and consequently, would be 
easier to establish from CMB observations. A program of study to relate the 
BipoSH brightness fluctuations in the baryon-photon fluid for different physical 
scenarios is currently underway. We have used our formalism to represent and match the well 
known case for SI violation in presence of a homogeneous magnetic field.
We illustrate how the angular correlations in such a case could be seeded by 
the dipole term of the generalized CMB brightness fluctuation and would have 
diagonal ($\ell' = \ell$) and off-diagonal ($\ell' = \ell \pm 2$) terms.

In summary, our work strongly motivates closer study
of all possible SI violating phenomena and scenarios in the simple
baryon-photon plasma, since these could potentially provide more
readily observable signature of SI violation in the universe.
It is also encouraging that it may have observational implications
in light of the recent WMAP-7 discovery \cite{bennett}. 
Since, the quadrupolar anisotropy anomaly with non-zero BipoSH coefficients related as $A_{\ell \ell}^{2M} \sim -2 A_{\ell-2 \ell}^{2M}$, 
rule out an origin in anisotropic power spectrum, this may well be related to SI violations in the CMB brightness fluctuations. 
This possibility is also bolstered by the fact that the non-SI effect peaks at acoustic $l\sim 200$ scales pointing to some non-trivial physics at 
the last scattering surface.
Extension of this formalism to CMB polarization should be readily possible. The formalism and the initial conclusions are 
important and timely in light of higher precision and resolution CMB anisotropy and polarization data expected in near future, in particular, 
from the ongoing Planck Surveyor CMB mission.

\appendix
\begin{widetext}
\vspace{3mm}
\section{\label{app0} Bipolar coefficients for SI deviations}

\noindent The BipoSH coefficients are defined in equation (\ref{biposh}). For a directional dependent primordial power spectrum as in equation (\ref{k-ps}), 
these Bipolar coefficients can be evaluated using the angular correlations in equation (\ref{k-dep}) in the following way
\ba
A^{J N}_{\ell \ell'} = \left(A^{J N}_{\ell \ell'}\right)_{\mathfrak{l}=0} &+& \sqrt{4 \pi} \int \frac{k^2 dk}{2 \pi^2} P_0(k)
\sum_{\mathfrak{l} \ell_1 \ell_2 L L'} \frac{\Pi_{\ell_1 \ell_3}}{\Pi_{\ell_2}} \beta_{\ell_1 \ell} \beta_{\ell_2 \ell'}^* \cg {\ell_2} {0} {\ell_1} {0} {\mathfrak{l}} {0}
\sum_{\mathfrak{m}} g_{\mathfrak{l} \mathfrak{m}}  \sum_{M M'} \tilde{\Delta}_{\ell_1 \ell} ^{L M}(k,\tau) \; [\tilde{\Delta}_{\ell_2 \ell'} ^{L' M'}(k,\tau)]^* \nonumber \\
& \times & \sum_{m_1 m_2 m m'} (-1)^{m'} \cg {L} {M} {\ell_1} {m_1} {\ell} {m} \cg {L'} {M'} {\ell_2} {m_2} {\ell'} {m'}
\cg {J} {N} {\ell} {m} {\ell'} {-m'} \cg {\ell_2} {m_2} {\ell_1} {m_1} {\mathfrak{l}} {\mathfrak{m}}.
\label{bip-co}
\ea
We use the following symmetry properties of the Clebsch-Gordan coefficients in equation (8.4.10) from \cite{varsha}
\ba
\cg {\ell_2} {m_2} {\ell_1} {m_1} {\mathfrak{l}} {\mathfrak{n}} &=& (-1)^{\mathfrak{l}+\ell_1-\ell_2} \cg {\ell_2} {-m_2} {\ell_1} {-m_1} {\mathfrak{l}} 
{-\mathfrak{m}} \nonumber \\
& = & (-1)^{\ell_1-\ell_2} \frac{\Pi_{\ell_2}}{\Pi_{\ell_1}} (-1)^{\mathfrak{m}} \cg {\ell_1} {m_1} {\ell_2} {m_2} {\mathfrak{l}} {-\mathfrak{m}}
\ea
and
\ba
\cg {J} {N} {\ell} {m} {\ell'} {-m'} &=& (-1)^{\ell'-m'} \frac{\Pi_{J}}{\Pi_{\ell}} \cg {\ell} {m} {\ell'} {m'} {J} {N} , 
\ea
where $\Pi_{\ell_1\ell_2 ...\ell_n} = [(2\ell_1 +1)(2\ell_2 +1)...(2\ell_n +1)]^{\frac{1}{2}}$ has been defined for convenient notational simplicity. \\
\noindent We use the formula for summation of the product of four Clebsch-Gordan coefficients given in equation (8.7.26) from \cite{varsha}
\ba
\sum_{m_1 m_2 m m'} \cg {L} {M} {\ell_1} {m_1} {\ell} {m} \cg {L'} {M'} {\ell_2} {m_2} {\ell'} {m'} \cg {\ell} {m} {\ell'} {m'} {J} {N}
\cg {\ell_1} {m_1} {\ell_2} {m_2} {\mathfrak{l}} {-\mathfrak{m}} = \sum_{\ell_3 m_3} \Pi_{\ell \ell_1 L' \ell_3} \cg {\ell_3} {m_3} {L'} {M'} {J} {N}
\cg {L} {M} {L'} {M'} {\ell_3} {m_3} \wninej{L}{\ell_1}{\ell}{L'}{\ell_2}{\ell'}{\ell_3}{\mathfrak{l}}{J} \; .
\ea
The Bipolar coefficients in equation (\ref{bip-co}) simplifies to
\ba
A^{J N}_{\ell \ell'} = \left(A^{J N}_{\ell \ell'}\right)_{\mathfrak{l}=0} &+& \sqrt{4 \pi} \int \frac{k^2 dk}{2 \pi^2} P_0(k)
\sum_{\mathfrak{l} \ell_1 \ell_2 L L'} \Pi_{\mathfrak{l} \ell_1  L L'} (-1)^{\mathfrak{l} + L'+ \ell' + \ell_1 - \ell_2} \: \beta_{\ell_1 \ell} \: \beta_{\ell_2 \ell'}^* \:
\cg {\ell_2} {0} {\ell_1} {0} {\mathfrak{l}} {0}  \nonumber \\
& \times & \sum_{\mathfrak{l}} g_{\mathfrak{l} \mathfrak{m}} \sum_{\ell_3 m_3} 
\wninej {J} {\ell_3} {\mathfrak{l}} {\ell} {L} {\ell_1} {\ell'}  {L'} {\ell_2} \:
\cg {J} {N} {\ell_3} {m_3} {\mathfrak{l}} {\mathfrak{m}} \left\{\tilde{\Delta}^L_{\ell_1 \ell} \otimes \tilde{\Delta}^{L'}_{\ell_2 \ell'}\right\}_{\ell_3 m_3}. 
\ea
$\{\tilde{\Delta}^L_{\ell_1 \ell} \otimes \tilde{\Delta}^{L'}_{\ell_1 \ell'}\}_{J N}$
are the Bipolar products in $L$ and $L'$ as defined in equation (\ref{new-biposh}). 

\vspace{5mm}
\noindent For statistically isotropic primordial perturbations, the angular correlation in equation (\ref{si-corr}) can be written in terms of Bipolar coefficients as
\ba         
(A^{J N}_{\ell \ell'})_{\mathfrak{l}=0} &=& 4 \pi \int \frac{k^2 dk}{2 \pi^2} P_0(k) \sum_{\ell_1 L L'} \beta_{\ell_1 \ell} \:\: \beta_{\ell_1 \ell'}^* 
\sum_{M M'} \tilde{\Delta}_{\ell_1 \ell} ^{L M}(k,\tau) \:\: [\tilde{\Delta}_{\ell_1 \ell'} ^{L' M'}(k,\tau)]^* \nonumber \\
& \times & \sum_{m_1 m m'} (-1)^{m'} \cg {J} {N} {\ell} {m} {\ell'} {-m'} \cg {L} {M} {\ell_1} {m_1} {\ell} {m} \cg {L'} {M'} {\ell_1} {m_1} {\ell'} {m'} .
\ea
The summation in the above equation can be simplified using equation (8.7.17) from \cite{varsha} as follows
\ba
\sum_{m_1 m m'} (-1)^{m'} \cg {J} {N} {\ell} {m} {\ell'} {-m'} \cg {L} {M} {\ell_1} {m_1} {\ell} {m} \cg {L'} {M'} {\ell_1} {m_1} {\ell'} {m'} 
& = & (-1)^{\ell_1+L+J} \: \Pi_{L' \mathcal{L''}} \cg {L} {M} {L'} {M'} {J} {N} \wsixj {\ell'} {\ell_1} {L'} {L} {J} {\ell}  \nonumber \\
& = & (-1)^{\ell_1+L+J} \: \Pi_{L' \mathcal{L''}} (-1)^{L'-M'} \: \frac{\Pi_L}{\Pi_{J}} \cg {J} {N} {L} {M} {L'} {-M'}
\wsixj{L}{J}{L'}{\ell'}{\ell_1}{\ell}  \nonumber \\
& = & (-1)^{\ell_1+L+L'+J} \: \Pi_{L L'} (-1)^{M'} \cg {J} {N} {L} {M} {L'} {-M'} \:
\wsixj{L}{J}{L'}{\ell'}{\ell_1}{\ell} \; .
\ea
Thus the BipoSH coefficient can be simplified to 
\ba
(A^{J N}_{\ell \ell'})_{\mathfrak{l}=0} &=& 4 \pi \int \frac{k^2 dk}{2 \pi^2} P_0(k) \sum_{\ell_1 L L'} (-1)^{\ell_1+L+L'+J} \: \Pi_{L L'} 
\beta_{\ell_1 \ell} \:\: \beta_{\ell_1 \ell'}^* \wsixj{L}{J}{L'}{\ell'}{\ell_1}{\ell}  \nonumber \\
& \times & \sum_{M M'} \tilde{\Delta}_{\ell_1 \ell} ^{L M}(k,\tau) \:\: [\tilde{\Delta}_{\ell_1 \ell'} ^{L' M'}(k,\tau)]^* (-1)^{M'} 
\cg {J} {N} {L} {M} {L'} {-M'} \nonumber \\
& = & 4 \pi \int \frac{k^2 dk}{2 \pi^2} P_0(k) \sum_{\ell_1 L L'} (-1)^{\ell_1+L+L'+J} \: \Pi_{L L'} 
\beta_{\ell_1 \ell} \:\: \beta_{\ell_1 \ell'}^* \wsixj{L}{J}{L'}{\ell'}{\ell_1}{\ell}  \nonumber \\
& \times & \left\{\tilde{\Delta}^L_{\ell_1 \ell} \otimes \tilde{\Delta}^{ L'}_{\ell_1 \ell'}\right\}_{J N} \; ,
\ea
where symmetry properties of the Clebsch-Gordan coefficients, equations (8.4.10) in \cite{varsha} has been used. 
\vspace{5mm}
\noindent To consider low Bipolar deviations from statistical isotropy i.e.  $(L , L', J << \ell , \ell', \ell_1)$, we use the asymptotic relation for 
Wigner-6j function as given in equation (9.9.1) from \cite{varsha} 
\ba
\wsixj {L} {J} {L'} {\ell'} {\ell_1} {\ell} & \approx & \frac{(-1)^{L+J+\ell'-\ell_1}}{\sqrt{2\ell_1} \; \Pi_{L'}} 
\cg {L'} {(\ell'-\ell_1)} {L} {(\ell-\ell_1)} {J} {(\ell'-\ell)} \nonumber \\
&\approx& \frac{(-1)^{J+\ell+\ell'}}{\sqrt{2\ell_1} \; \Pi_{J}} \cg {J} {(\ell-\ell')} {L} {(\ell-\ell_1)} {L'} {(\ell_1 -\ell')}.
\ea
Using the above relation, the asymptotic limit to the BipoSH coefficients turn out to be
\ba
(A^{J N}_{\ell \ell'})_{\mathfrak{l}=0} & \approx &  \!4 \pi  \! \int \frac{k^2 dk}{2 \pi^2} P_0(k) \! \sum_{\ell_1 L L'} (-1)^{\ell + \ell'+ \ell_1+L+L'} \!
\frac{\Pi_{L L'}}{\sqrt{2\ell_1} \: \Pi_{J}} \: \beta_{\ell_1 \ell} \: \beta_{\ell_1 \ell'}^* \:
\cg {J} {(\ell-\ell')} {L} {(\ell-\ell_1)} {L'} {(\ell_1-\ell')} 
\left\{\tilde{\Delta}^L_{\ell_1 \ell} \otimes \tilde{\Delta}^{ L'}_{\ell_1 \ell'}\right\}_{J N} \!.
\ea
\section{\label{appA} Generalized evolution equation for statistical isotropy breakdown}

\noindent Starting with equation (\ref{evolgen}), putting $P_\ell(\hat{k} \cdot \hat{n}) = \frac{4\pi}{2\ell +1} \sum_m Y_{\ell m}^*(\hat{k}) \: Y_{\ell m}(\hat{n})$
and evaluating the double integral using the equation
\be
\int d\Omega_{\hat{n}}  Y_{\ell m}(\hat{n}) \; Y_{\ell_1 m_1}(\hat{n}) Y_{\ell_2 m_2}(\hat{n})^*
 =\frac{1}{\sqrt{4 \pi}} \; \frac{\Pi_{\ell \ell_1}}{\Pi_{\ell_2}} \mathcal{C}^{\ell_2 0}_{\ell 0 \ell_1 0} \; \mathcal{C}^{\ell_2 m_2}_{\ell m \ell_1 m_1}, 
\ee
the most general evolution equation becomes
\be
\tilde{\Delta}_{\ell_1 \ell_2} ^{L M}(k,\tau) = \frac{1}{\beta_{\ell_1 \ell_2}} \sum_\ell (-i)^\ell \;\Pi_\ell^2 \; j_\ell(k \Delta \tau)
\sum_{\ell_3 \ell_4 L' M'} \!\!\!\!\!\!  \beta_{\ell_3 \ell_4} \tilde{\Delta}_{\ell_3 \ell_4} ^{L' M'}(k,\tau_s)
\frac{\Pi_{\ell_1 \ell_4}}{\Pi_{\ell_2 \ell_3}}  \mathcal{C}_g \; ,
\label{app-gen-evol}
\ee
where
\be
\mathcal{C}_g = \mathcal{C}^{\ell_3 0}_{\ell 0 \ell_1 0} \; \mathcal{C}^{\ell_2 0}_{\ell 0 \ell_4 0}
\sum_{m_1 m_2} \mathcal{C}^{L M}_{\ell_1 m_1 \ell_2 m_2} \sum_{m_3 m_4} \mathcal{C}^{L' M'}_{\ell_3 m_3 \ell_4 m_4}
\sum_m \mathcal{C}^{\ell_3 m_3}_{\ell m \ell_1 m_1} \; \mathcal{C}^{\ell_2 m_2}_{\ell m \ell_4 m_4}
\label{clebsch}
.\ee
Using the summation formula for four Clebsch-Gordon coefficients given by equation (9.1.8) from \cite{varsha}, 
\be
\sum_{m m_1 m_2 m_3 m_4} \cg {L'} {M'} {\ell_3} {m_3} {\ell_4} {m_4} \; \cg {\ell_3} {m_3} {\ell_1} {m_1} {\ell} {m}\; \cg {L} {M} {\ell_1} {m_1} {\ell_2} {m_2}
\cg {\ell_2} {m_2} {\ell} {m} {\ell_4} {m_4}
=\delta_{L' L} \delta_{M' M} (-1)^{\ell_1+ \ell+ \ell_4+ L'}
\Pi_{\ell_2 \ell_3} \wsixj {\ell_1} {\ell} {\ell_3} {\ell_4} {L'} {\ell_2} \; ,
\ee
equation (\ref{app-gen-evol}) simplifies to
\ba
\tilde{\Delta}_{\ell_1 \ell_2} ^{L M}(k,\tau) &=& \sum_{\ell \ell_3 \ell_4} (-i)^{\ell} \; \Pi_{\ell \ell \ell_1 \ell_4}
\frac{\beta_{\ell_3 \ell_4}}{\beta_{\ell_1 \ell_2}} \; (-1)^{\ell_3 + \ell_4 + L} \; j_\ell(k \Delta \tau)
\tilde{\Delta}_{\ell_3 \ell_4} ^{L M}(k,\tau_s) \; \cg {\ell_3} 0 {\ell} 0 {\ell_1} 0 \;
\cg {\ell_2} 0 {\ell} 0 {\ell_4} 0 \; \wsixj {\ell_1} {\ell}  {\ell_3} {\ell_4} L {\ell_2} \nonumber \\
&=&\sum_{\ell \ell_3 \ell_4} (-i)^{\ell} \; \Pi_{\ell \ell \ell_1 \ell_4}
\frac{\beta_{\ell_3 \ell_4}}{\beta_{\ell_1 \ell_2}} \; (-1)^{\ell_3 + \ell_4 + L} \; j_\ell(k \Delta \tau)
\tilde{\Delta}_{\ell_3 \ell_4} ^{L M}(k,\tau_s) \; \cg {\ell_3} 0 {\ell} 0 {\ell_1} 0 \;
\cg {\ell_2} 0 {\ell} 0 {\ell_4} 0 \; \wsixj {\ell_1} L {\ell_2} {\ell_4} {\ell} {\ell_3}.
\label{short-evol}
\ea
Substituting the multipole dependent coefficients as
\be
C(L,\ell,\ell_1,\ell_2) = (-i)^{\ell} \; \Pi_{\ell \ell \ell_1 \ell_4}
\frac{\beta_{\ell_3 \ell_4}}{\beta_{\ell_1 \ell_2}} \; (-1)^{\ell_3 + \ell_4 + L} \cg {\ell_3} 0 {\ell} 0 {\ell_1} 0 \;
\cg {\ell_2} 0 {\ell} 0 {\ell_4} 0 \; \wsixj {\ell_1} L {\ell_2} {\ell_4} {\ell} {\ell_3} \;,
\ee
the generalized evolution equation for deviations of statistical isotropy in the CMB brightness fluctuation reduces to 
\be
\tilde{\Delta}_{\ell_1\ell_2}^{L M}(k,\tau) = \sum_{\ell \ell_3 \ell_4} C(L, \ell,\ell_1,\ell_2) \; j_\ell(k \Delta \tau) \; 
\tilde{\Delta}_{\ell_3 \ell_4} ^{L M}(k,\tau_s) \; .
\ee
\vspace{3mm}
\section{\label{appB} Diagonal terms of the asymptotic moments of CMB brightness fluctuation}

\noindent In equation (\ref{asymp}), taking $e=0$ i.e. $\ell = R$ and putting $\ell_1 = R + \ell_4 $ \& $\ell_2 = R + \ell_3$, we can write equation (\ref{sixj}) as
\be
\wsixj {\ell_4} L {\ell_3} {\ell + \ell_4} {\ell} {\ell + \ell_3} \approx \frac{(-1)^{L}}{\sqrt{2\ell}\:\Pi_{\ell_3}}
\cg {\ell_3} {\ell_4} {\ell_4} {\ell_3} L {(\ell_4-\ell_3)}
.\ee
\noindent From the properties of the Clebsch-Gordon coefficient, we get $\ell_3 = \ell_4$ and hence $\ell_1 = \ell_2$. 
Thus in equation (\ref{simevol}) substituting $\ell_3 = \ell_1 -\ell$, we get the factor
\be
C_1(L, \ell, \ell_1) = (-i)^{\ell} \; \Pi_{\ell}^2 \; \Pi_{\ell_1} \; \Pi_{\ell_1 - \ell} \frac{\beta_{\ell_1 - \ell, \ell_1 - \ell}}{\beta_{\ell_1,\ell_1}} 
(-1)^L \; \cg {(\ell_1-\ell)} 0 {\ell} 0 {\ell_1} 0 \; \cg {\ell_1} 0 {\ell} 0 {(\ell_1-\ell)} 0  \frac{(-1)^{L}}{\sqrt{2\ell} \, \Pi_{\ell_1-\ell}}
\cg {(\ell_1-\ell)} {(\ell_1-\ell)} {(\ell_1-\ell)} {(\ell_1-\ell)} L 0 \;\; .
\ee
Now using equation (8.4.10) from \cite{varsha}, 
\ba
\cg {(\ell_1-\ell)} 0 {\ell} 0 {\ell_1} 0 &=& (-1)^{\ell+\ell_1-(\ell_1-\ell)} \cg {(\ell_1-\ell)} 0 {\ell_1} 0 {\ell} 0 = \cg {(\ell_1-\ell)} 0 {\ell_1} 0 {\ell} 0 , \nonumber \\
\cg {\ell_1} 0 {\ell} 0 {(\ell_1-\ell)} 0 &=& (-1)^{\ell} \frac{\Pi_{\ell_1}}{\Pi_{\ell_1 - \ell}} \cg {(\ell_1-\ell)} 0 {\ell_1} 0 {\ell} 0 .
\label{sym}
\ea
\noindent  Using equations (8.5.34) and (8.5.42) from \cite{varsha}, we get
\be
\left[ \cg {(\ell_1-\ell)} 0 {\ell_1} 0 {\ell} 0 \right] ^2 = \left[ \frac{\ell_1 !}{\ell! (\ell_1 -\ell)!} \right] ^2
\frac{(2\ell)!(2\ell_1-2\ell+1)!}{(2\ell_1+1)!} ,
\ee
and
\be
\cg {(\ell_1-\ell)} {(\ell_1-\ell)} {(\ell_1-\ell)} {(\ell_1-\ell)} L 0 = (2\ell_1-2\ell)! \sqrt{\frac{2\ell_1-2\ell +1}{(2\ell_1-2\ell -L)!(2\ell_1-2\ell+L+1)!}} \;\; .
\ee
Thus
\be
C_1(L,\ell, \ell_1) =  \frac{\Pi_{\ell}^2 \Pi_{\ell_1}\Pi_{\ell_3}}{\sqrt{2\ell}}
\left[ \frac{\ell_1!}{\ell! \ell_3!} \right] ^2
\frac{(2\ell)!(2\ell_3+1)!}{(2\ell_1+1)!} \;\;
\sqrt{\frac{(2\ell_3)!}{(2\ell_3 -L)!} \frac{(2\ell_3)!}{(2\ell_3+L+1)!}} \;\; \delta_{\ell_3,\ell_1-\ell} \;\; .
\ee
Substituting this in equation (\ref{simevol}), the evolution equation becomes
\be
\tilde{\Delta}_{\ell_1, \ell_1} ^{L M}(k,\tau) = \sum_{\ell} C_1(L, \ell,\ell_1) \; j_\ell(k \Delta \tau) \;
\tilde{\Delta}_{\ell_1-\ell ,\; \ell_1-\ell} ^{L M}(k,\tau_s) \;\; ,
\ee
where $\ell, \ell_1 >> |\ell_1 - \ell|$.
\vspace{3mm}
\section{\label{appC} Off-diagonal terms of the asymptotic moments of CMB brightness fluctuation}

\noindent In equation (\ref{asymp}), with $e=0$ i.e. $\ell = R$
\be
\wsixj {\ell_4} L {\ell_3} {\ell_1} {\ell} {\ell_2} \approx \frac{(-1)^{\ell_4+L+\ell_1 - \ell}}{\sqrt{2\ell} \:\Pi_{\ell_3}}
\cg {\ell_3} {(\ell_1-\ell)} {\ell_4} {(\ell_2-\ell)} L {(\ell_1-\ell_2)}
.\ee
\noindent From conditions of Clebsch-Gordon coefficients
\ba
|\ell_1 -\ell| \leq \ell_3 \nonumber \\
|\ell_2 -\ell| \leq \ell_4 \nonumber \\
|\ell_1 -\ell_2| \leq L \nonumber
.\ea
\noindent For off-diagonal terms of $\tilde{\Delta}_{\ell_1, \ell_2} ^{L M}$ we consider the minimum values of
$\ell_3$, $\ell_4$ \&  $L$ with neither equal to 0.
\ba
\ell_1 -\ell = \ell_3 \nonumber \\
\ell_2 -\ell = \ell_4 \nonumber \\
\ell_1 -\ell_2 =  L  \nonumber .
\ea
Thus in equation (\ref{simevol}) we would get the factor
\be
C_2(\ell, \ell_1, \ell_2) = (-i)^{\ell} \Pi_{\ell}^2 \Pi_{\ell_1} \Pi_{\ell_2-\ell} \frac{\beta_{\ell_1-\ell,\ell_2-\ell}}{\beta_{\ell_1 \ell_2}}
\frac{(-1)^{L+\ell_1+\ell_2}}{\sqrt{2\ell}\:\Pi_{\ell_1 -\ell}} \; \; \cg {(\ell_1 -\ell)} 0 {\ell} 0 {\ell_1} 0
\cg {\ell_2} 0 {\ell} 0 {(\ell_2 -\ell)} 0 \;\;
\cg {(\ell_1 -\ell)} {(\ell_1 -\ell)} {(\ell_2-\ell)} {(\ell_2-\ell) \;}  L L \;\; .
\ee
\vspace{5mm}
\noindent Now using the symmetries of Clebsch-Gordon coefficients given by equation (8.5.34) from \cite{varsha}, we get
\ba
\cg {(\ell_1 -\ell)} 0 {\ell} 0 {\ell_1} 0 &=& \cg {(\ell_1 - \ell)} 0 {\ell_1} 0 {\ell} 0  \nonumber \\
&=&  \frac{(-1)^{\ell} \ell_1 !}{\ell! (\ell_1 - \ell)!}
\left[\frac{(2\ell)!(2\ell_1 -2 \ell + 1)!}{(2\ell_1+1)!} \right] ^{\frac{1}{2}}
\label{sym1}
.\ea
\noindent Similarly
\ba
\cg {\ell_2} 0 {\ell} 0 {(\ell_2-\ell)} 0 &=& (-1)^{\ell} \frac{\Pi_{\ell_2}}{\Pi_{\ell_2-\ell}}
\cg {(\ell_2 - \ell)} 0 {\ell_2} 0 {\ell} 0  \nonumber \\
&=& \frac{\Pi_{\ell_2}}{\Pi_{\ell_2-\ell}} \frac{\ell_2 !}{\ell! (\ell_2 - \ell)!}
\left[\frac{(2\ell)!(2\ell_2 -2 \ell + 1)!}{(2\ell_2+1)!} \right] ^{\frac{1}{2}}
\label{sym2}
.\ea
\vspace{3mm}
\noindent Using  equation (8.5.37) from \cite{varsha} we get
\ba
\cg {(\ell_1-\ell)} {(\ell_1-\ell)} {(\ell_2-\ell)} {(\ell_2-\ell)} L L &=& \delta_{\ell_2-\ell+L, \ell_1 -\ell} \nonumber \\
&=& \delta_{\ell_2+ L, \ell_1}
.\ea
Thus
\be
C_2(\ell, \ell_1, \ell_2) =  \frac{(-1)^{\frac{\ell_1 + \ell_2}{2}}}{\sqrt{2\ell}} \frac{(2\ell)! \ell_1 ! \ell_2 ! }{(\ell!)^2 (\ell_1 - \ell)! (\ell_2 - \ell)!}
\left[\frac{(2\ell_1 -2 \ell + 1)! (2\ell_2 -2 \ell + 1)!}{(2\ell_1+1)! (2\ell_2+1)!} \right] ^{\frac{1}{2}}
\frac{\Pi_{\ell \ell \ell_1 \ell_2}}{\Pi_{\ell_1-\ell}} \: \sqrt{\frac{\ell_1+\ell_2-2\ell+1}{\ell_1+\ell_2+1}}
.\ee
\noindent Substituting this in equation (\ref{simevol}), the evolution equation becomes
\be
\tilde{\Delta}_{\ell_1, \ell_2}^{L M}(k,\tau) = \sum_{\ell} C_2(\ell,\ell_1,\ell_2) \; j_{\ell}(k\Delta \tau) 
\; \tilde{\Delta}_{\ell_1 -\ell, \ell_2-\ell} ^{L M}(k,\tau_s) \; \delta_{L,\ell_1 - \ell_2} \; ,
\ee
where $\ell,\; \ell_n >> |\ell_n -\ell|,\; L$ with $n = 1,2$.
\end{widetext}
\bibliographystyle{apsrev}
\bibliography{si}		
\end{document}